\documentclass[aps,prl,groupedaddress,twocolumn]{revtex4}
\usepackage{graphicx}
\usepackage{amsmath}
\usepackage{color}
\usepackage[sort&compress]{natbib}

\begin{document}

\title{Toroidal Optical Activity}

\author{T. A. Raybould}
\affiliation{Optoelectronics Research Centre, University of
Southampton, SO17 1BJ, United Kingdom} \email{T.A.Raybould@soton.ac.uk}

\author{V. A. Fedotov}
\affiliation{Optoelectronics Research Centre, University of
Southampton, SO17 1BJ, United Kingdom}

\author{N. Papasimakis}
\affiliation{Optoelectronics Research Centre, University of
Southampton, SO17 1BJ, United Kingdom}

\author{I. Kuprov}
\affiliation{School of Chemistry, University of
Southampton, SO17 1BJ, United Kingdom}

\author{I. Youngs}
\affiliation{DSTL, Salisbury, United Kingdom}

\author{N. I. Zheludev}
\affiliation{Optoelectronics Research Centre, University of
Southampton, SO17 1BJ, United Kingdom}

\author{W. T. Chen}
\affiliation{Department of Physics, National Taiwan University, Taipei, 10617, Taiwan}

\author{D. P. Tsai}
\affiliation{Department of Physics, National Taiwan University, Taipei, 10617, Taiwan}

\date{\today}

\begin{abstract}
Optical activity is ubiquitous across natural and artificial media and is conventionally understood in terms of
scattering from electric and magnetic moments. Here we demonstrate experimentally and confirm numerically a type of
optical activity that cannot be attributed to electric and magnetic multipoles. We show that our observations can
only be accounted for by the inclusion of the toroidal dipole moment, the first term of the recently established
peculiar family of toroidal multipoles.
\end{abstract}

\maketitle

Chirality, i.e. the property by which a pattern cannot be superimposed with its mirror image \cite{kelvin1904}, is
a basic characteristic of structured matter at all scales, from the cosmological to the molecular. It enters
fundamental questions, such as the homochirality observed in life \cite{saito2013} and the parity violation within
the Standard Model of physics \cite{lee1956,wang2014}. An ubiquitous manifestation of chirality is optical
activity, an optical effect characterized by the rotation of the polarization plane of linearly polarized light
propagating through a chiral medium, and preferential absorption of circularly polarized light of a particular
handedness (circular dichroism) if the medium is dissipative. Observed more than 150 years ago by Louis Pasteur on
solutions of tartaric acid \cite{pasteur1848}, optical activity occupies a central position in the diagnostic
methodology of many scientific disciplines \cite{Barron2004}, where it provides means of obtaining information about
the microscopic structure and electromagnetic excitations of a medium from its far-field scattering properties. An
examination of optical activity is invaluable, for example, in the biosciences where it enables deduction of the
conformation of proteins in a dynamic fashion \cite{CDproteins1}, or in the study of magnetic phenomena in strongly
correlated electronic systems \cite{rasing2010}.

Optical activity is typically described within the dipole approximation of the dynamic multipole expansion \cite
{Barron2004}, where the interaction of an incident wave with collinear electric and magnetic dipoles induced in the
medium affects the polarization state of the wave upon transmission. Certain combinations of electric dipole and
other higher-order multipoles (most notably electrical quadrupole) can also contribute to optical activity \cite
{shellman1968,Raab2005}, although such contributions are usually considered negligible, especially in isotropic
media \cite{nakano1969,dunn1971}. It was suggested recently \cite{Papasimakis2009} that optical activity could
arise even in the absence of electric dipolar response due to the excitation of toroidal multipoles, an elusive
part of the dynamic multipole response \cite{Dubovik1990,Radescu2002}. In particular, it was anticipated that
toroidal dipole should contribute to optical activity in the same manner as electric dipole \cite{Papasimakis2009}
since the two have the same angular momentum and parity properties \cite{Radescu2002,Mexican}.

In this Letter we observe experimentally and confirm numerically that the toroidal dipolar excitation can in fact
provide the dominant contribution to optical activity, assuming the role of the electric dipole. Using the
metamaterial approach, we have been able to enhance in a resonant fashion the microscopic toroidal and electric
quadrupole responses to circularly polarized radiation and show that this previously unexplored and rather exotic
combination of dynamically induced multipoles leads to strong circular dichroism. Our results challenge the common
interpretation of optical activity based on the electric dipole-magnetic dipole approximation and enhance its
capabilities as a diagnostic tool.

\begin{figure}[t]
\includegraphics[width=86mm]{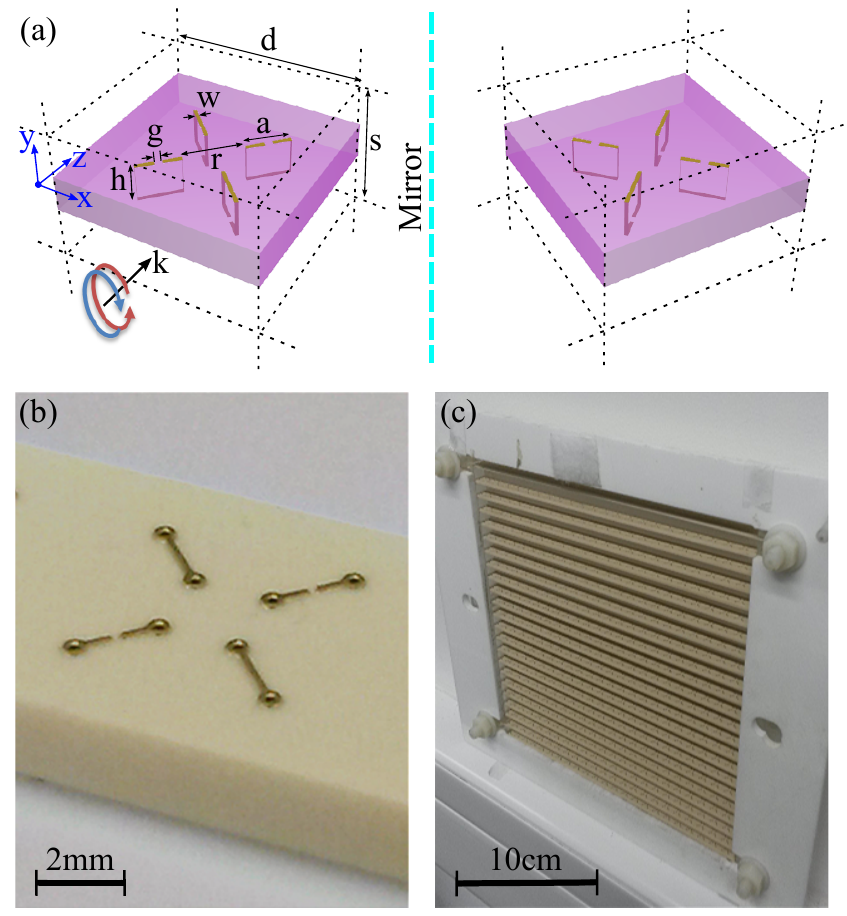} \caption{(a) A schematic of the toroidal metamolecule and its
orientatation with respect to the metamaterial crystal lattice (indicated by dashed lines). The dimensions are
$d=8$mm, $s=7.5$mm, $h=1.5$mm, $r=2.44$mm, $g=w=0.15$mm and $a=1.8$mm. The metamaterial slab is formed by
translating the unit cell along the $x$ and $y$ axes, which imposes structural chirality on the sample. RCP (red)
and LCP (blue) light propagates along the $z$--axis of the metamaterial. Mirror image shows the enantiomeric form
of the metamaterial crystal. (b) Close-up photograph of the fabricated toroidal metamolecule. (d) Metamaterial
sample after assembly. }
\label{tDipMM}
\end{figure}

The metamaterial used for this study has been derived from the structure previously shown to support a strong
toroidal dipolar response under linearly polarized excitation \cite{Kaelberer2010}. Its metamolecule consists of a
cluster of four rectangular metallic wire loops embedded in a dielectric slab [see Fig. \ref{tDipMM}(a)]. The loops
are located in two mutually orthogonal planes, whose intersection gives the axis of the metamolecule parallel to
$y$--axis. Each loop is split in the centre of either its top or bottom side depending on which of the two
orthogonal planes it is located in. The metamaterial is formed by translating the metamolecule along the $x$ and
$y$ axes, resulting in a one-metamolecule thick crystal slab with a rectangular unit cell, as shown in Fig.
\ref{tDipMM}(a). Although the individual metamolecules are intrinsically achiral, the resulting metamaterial
crystal is chiral due to the regular arrangement of its metamolecules ('structural' chirality \cite{Zheludev2003}) and is
available in two enantiomeric forms interconnected by a mirror reflection [see Fig. 1(a)].

The metamaterial sample was fabricated by etching a 35$\mu$m thick copper foil on both sides of a low-loss
dielectric PCB laminate Rogers TMM(R) 3 using high-resolution photolithography [see Fig. 1(b)]. The top and bottom
patterns were electrically connected through narrow electroplated holes. All copper tracks were coated with a 2$\mu$m
thick layer of gold to prevent oxidization and staining of copper, as well as to reduce Ohmic losses. The assembled
metamaterial sample consisted of $24 \times 24$ metamolecules. Its transmission response was experimentally
characterized at normal incidence in a microwave anechoic chamber at frequencies in the range 12.5-16.5 GHz using
broadband linearly polarized horn antennas equiped with collimating lenses and a vector network analyzer. Our
measurements yielded a complex transmission matrix $t_{ij}$ ($i, j=x, y$), which fully describes the metamaterial
transmission in the linear polarization basis. To obtain the transmission matrix in the circular basis we used the
following well-known relations:

\begin{multline}
\hspace{3cm} \begin{bmatrix}
t_{++} & t_{+-}\\
t_{-+} & t_{--}
\end{bmatrix}= \\
\frac{1}{2}\begin{bmatrix}
 t_{xx}+t_{yy}+i  \left ( t_{xy}-t_{yx} \right ) & t_{xx}-t_{yy}-i  \left ( t_{xy}+t_{yx} \right ) \\
t_{xx}-t_{yy}+i  \left ( t_{xy}+t_{yx} \right ) &  t_{xx}+t_{yy}-i  \left ( t_{xy}-t_{yx} \right )
\end{bmatrix},
\label{t_conv}
\end{multline}

where the subscripts $+$ and $-$ denote right (RCP) and left (LCP) circular polarizations.

\begin{figure}
\includegraphics[width=86mm]{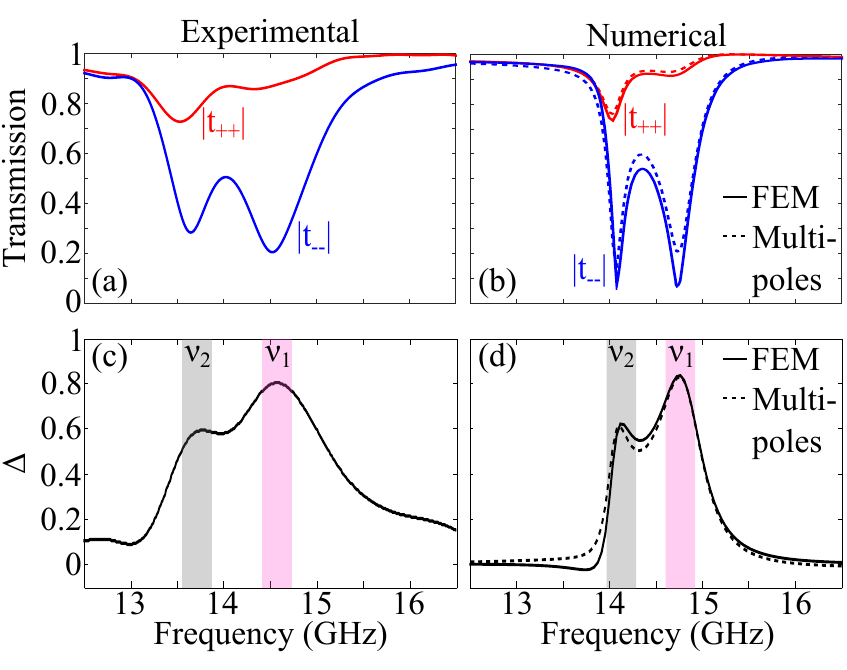}
\caption{Measured (left) and calculated (right) circular transmission response of the toroidal metamaterial. Panels
(a) and (b) show the amplitudes of the direct circular transmission $t_{++}$ and $t_{--}$, while panels (c) and (d)
-- circular dichroism $\Delta$. Results calculated using two different methods -- finite-element simulation and
multipole decomposition of currents are shown by solid and dashed curves, respectively. Purple and grey bands mark
the resonances of corresponding conventional and toroidal circular dichroism.}
\label{trans_dichro}
\end{figure}

As evident from Fig. \ref{trans_dichro}(a), the diagonal elements of the resulting matrix, $t_{++}$ and $t_{--}$,
do not match indicating the presence of optical activity in the response of the metamaterial. The difference
between the transmission intensities $\Delta=\left |t_{++}\right |^2 - \left |t_{--}\right |^2$ is a measure of
structure's circular dichroism and it is particularly strong within two resonant bands located at $\nu_1=14.6$ and
$\nu_2=13.6$ GHz, where the dichroism reaches 60\% and 80\% respectively [see Fig. \ref{trans_dichro}(b)].

In order to determine the cause of optical activity at these two resonances we conducted a full-wave simulation of
the metamaterial response using a 3D finite element solver (COMSOL 3.5a). The metamaterial array was modelled by placing the toroidal metamolecule in a rectangular unit cell with periodic boundary conditions imposed along $x$ and $y$ axes [see Fig. 1(a)]. The wires of the metallic loops were assumed to be infinitely thin strips of perfect electric conductors. Permittivity of the dielectric slab was set to $\epsilon= 3.45-i0.007$, which corresponds to Rogers TMM(R) 3 dielectric laminate used in the fabrication of the metamaterial sample. This model allowed us to accurately reproduce the main features of the metamaterial response, yielding a good agreement between the calculated and measured far-field transmission spectra (solid curves in Fig. \ref{trans_dichro}). The slight discrepancy is attributed to fabrication tolerances and uncertainty in the
dielectric constant of the PCB laminate.

The microscopic origin of the effect was analysed in terms of multipolar charge-current excitations supported by
the metamolecules (analogous to multipolar transitions in conventional molecules). The multipole moments were
computed based on spatial distributions of the induced conduction and displacement current densities extracted from
our model, as it was previously described in \cite{Kaelberer2010}. To simplify the analysis the expansion was
truncated at the octopole order for electric and magnetic multipoles, and at the quadrupole order for toroidal
multipoles. Following the recently proposed methodology \cite{Savinov2014} we can directly link the multipolar
excitations of individual metamolecules to the far-field response of the entire metamaterial array. An adequate
accuracy of this approach is demonstrated by Figs. \ref{trans_dichro}(c) and \ref{trans_dichro}(d) where we compare
the metamaterial transmission calculated through the multipole decomposition (dashed curves) with that obtained
directly from the simulations (solid curves). The small quantitative mismatch here is the result of limited spatial
resolution of extracting the current data rather than truncation of the multipole series.

\begin{figure}
\includegraphics[width=86mm]{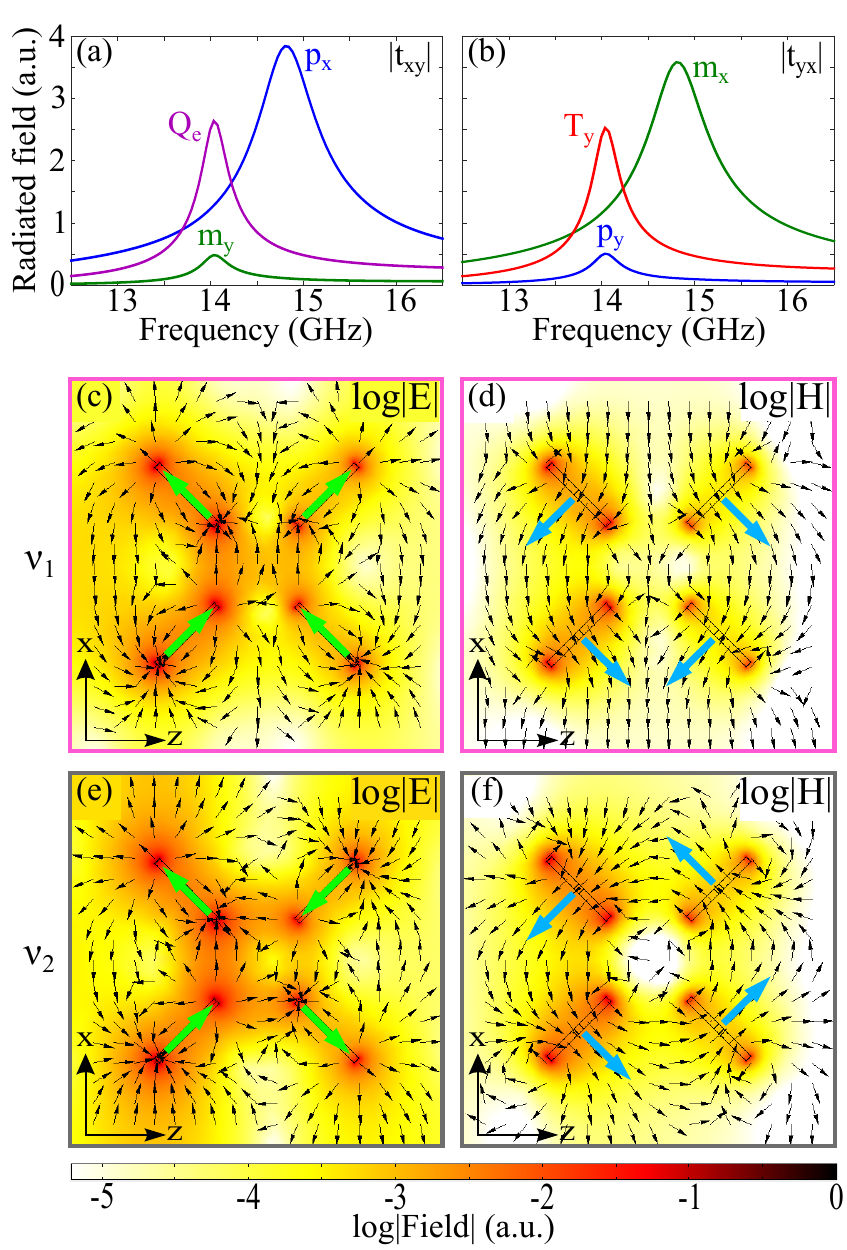}
\caption{Amplitudes of the fields radiated by the four most dominant multipole components (electric dipole $p$,
magnetic dipole $m$, toroidal dipole $T$ and electric quadrupole $Q_e$) that contribute to polarization conversion
response of the metamaterial $t_{xy}$ (a) and $t_{yx}$ (b). (c)-(f) give the field distributions ($log|E|$ and
$log|H|$) around a metamolecule when excited by LCP light at $\nu_1$ (c)-(d) and $\nu_2$ (e)-(f). The black arrows
give the corresponding $E$ and $H$ field direction. The green (blue) arrows show orientation of the excited
electric (magnetic) dipoles.}
\label{fieldamp}
\end{figure}

For the purpose of the analysis it is sufficient to perform the multipole decomposition of
the metamaterial's polarization conversion response, $t_{xy}$ and $t_{yx}$, since these can be regarded as a
manifestation of the observed circular dichroism in the linear basis. Indeed, according to Eq. (1) the difference
between the diagonal elements of the circular transmission matrix exists only for non-zero linear polarization
conversion terms, where $\left | t_{xy} \right | = \left | t_{yx} \right |$ as dictated by the reciprocity and
structural symmetry of the metamaterial with respect to the propagation direction ($z$-axis). Figures \ref
{fieldamp}(a) and \ref{fieldamp}(b) show the results of the multipole decomposition where we retained only six
multipole components for clarity, namely $x$-components of electric $\mathbf{p}$ and magnetic $\mathbf{m}$ dipoles
at $\nu_1$, and electric quadrupole $\mathbf{Q_{e}}$ and $y$-components of electric, magnetic and toroidal $
\mathbf{T}$ dipoles at $\nu_2$ (other multipole components are at least one order of magnitude smaller than the
dominant ones):

\vspace{0.2cm}

$\mathbf{p}=\frac{1}{i\omega } \int \mathbf{j}d^3r$
\vspace{0.2cm}

$\mathbf{m}=\frac{1}{2c}\int \left ( \mathbf{r}\mathbf{\times}\mathbf{j}  \right )d^3r$
 \vspace{0.2cm}

$\mathbf{T}=\frac{1}{10c}\int \left [ \left ( \mathbf{r\cdot j} \right )\mathbf{r} -2r^2\mathbf{j}\right ]d^3r$
\vspace{0.2cm}

 $Q_{\alpha \beta} =\frac{1}{2i\omega }\int \left [ r_\alpha j_\beta+r_\beta j_\alpha-\frac{2}{3}\delta_{\alpha
\beta}\left ( \mathbf{r\cdot j} \right ) \right ]d^3r$
\vspace{0.2cm}

where $\mathbf{j}$ is the induced current density.

In agreement with the modelled circular dichroism data [Fig. 2(d)], the resonances in the multipole scattering
occur at frequencies $\nu_1$ and $\nu_2$. Figs. 3(a) and 3(b) indicate a symmetry
to the linear conversion response though the multipoles contributing to the same resonant features in $t_{xy}$ and
$t_{yx}$ are different. This means that the effect of circular dichroism is controlled by essentially pairs of
different multipolar excitations. In particular, at $\nu_1$ the dominant pair comprises electric and magnetic
dipoles with collinear moments -- $p_{x}$ and $m_{x}$. Such a combination corresponds to the textbook optical
activity and hence is considered as a reference case in our further analysis below. At $\nu_2$, however, the
contributions of $p_{x}$ and $m_{x}$ become only secondary, while the circular dichroism appears to be underpinned
by previously unexplored and rather exotic combination of the toroidal dipole and electric quadrupole.

The presence of these multipole pairs can be visually detected in the distributions of electric and magnetic fields
inside the unit cell arising under circularly polarized excitation [see Figs. 3(c) - (f)]. At $\nu_1$ the field
lines of both electric and magnetic fields are seen to stream parallel to $x$-axis, which clearly correspond to
electric and magnetic dipolar excitations of the metamolecule with the net dipole moments being collinear and
oriented along $x$-axis [Figs. 3(c) and (d)]. At $\nu_2$ magnetic field is seen to confine within a well-defined
ring-like area where the field lines form a vortex threading all loops of the metamolecule [Fig. 3(e)]. Such a
magnetic field configuration is unique to the toroidal excitation with the net dipole moment aligned parallel to
the axis of the metamolecule (i.e. $y$-axis). The distribution of electric field features a pattern similar to that
in Fig. 3(c) but the field lines on the opposite sides of the metamolecule stream anti-parallel to each other [Fig.
3(f)], which is typical for an electric quadrupole excitation.

\begin{figure}[t]
\includegraphics[width=86mm]{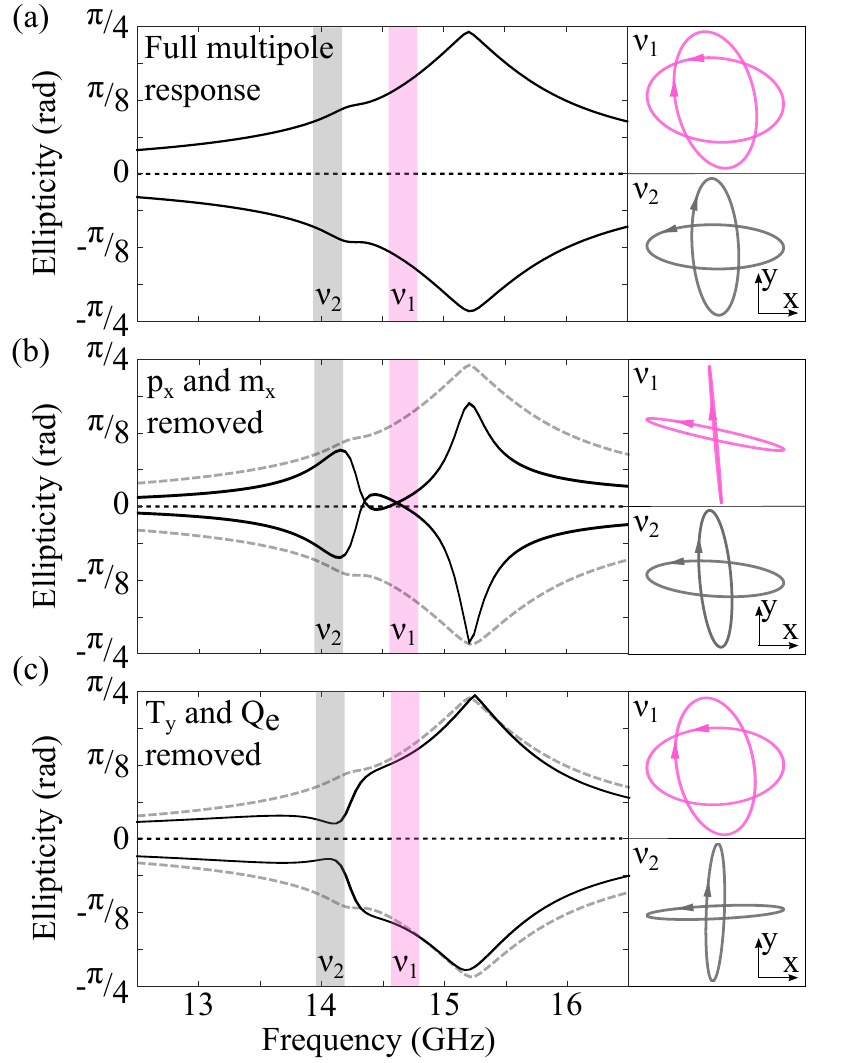}
\caption{Ellipticity angles of the metamaterial eigenstates (left) and associated polarization ellipses at $\nu_1$
and $\nu_2$ (right), calculated based on the multipole expansion of microscopic charge-current excitations
supported by toroidal metamolecules. Panel (a) shows the results obtained using the complete multipole set. Panels
(b) and (c) show the results corresponding to pairs $p_{x}-m_{x}$ and $T_{y}-Q_{e}$ being removed from the
microscopic response of the metamaterial (solid curves) and compared to the complete multipole response from the
upper panel (dashed curves).}
\label{ellipticities}
\end{figure}

Our conclusions regarding the nature of the observed circular dichroism can be further reinforced by examining the
role of the identified multipole pairs in shaping the metamaterial's polarization eigenstates. The eigenstates were
computed by solving the characteristic equation of the transmission matrix $\mathbf{t}$, which was reconstructed
from the multipole data. The polarization eigenstates were characterised in terms of their ellipticity angle $
\epsilon$ using the following relation:

\begin{equation}
\epsilon =  \frac{1}{2}arcsin\left ( \frac{\left | \mathbf{e_+} \right |^2-\left | \mathbf{e_-} \right |^2}{\left |
\mathbf{e_+} \right |^2+\left | \mathbf{e_-} \right |^2} \right ),
\label{ellipt}
\end{equation}

where $\mathbf{e_+}$ and $\mathbf{e_-}$ are eigenvectors of $\mathbf{t}$ in circular basis, and $\left | \epsilon
\right |$ takes values between $\pi/4$ (circular polarization) and 0 (linear polarization). A positive ellipticity
angle defines right handed rotation, while negative -- left handed rotation. The two limiting cases, $\epsilon =
\pm\pi/4$ and $\epsilon = 0$, represent polarization eigenstates of optically active (chiral) isotropic media and
linearly birefringent (achiral) anisotropic media, respectively. The calculated eigenstates of the metamaterial
correspond to the intermediate case of counter rotating elliptical polarizations with $\epsilon \approx \pm\pi/8$
near the resonances, where the ellipticity arises due to anisotropy in the metamaterial structure [see Fig. \ref
{ellipticities}(a)].

The roles of the multipoles are assessed by removing their contributions in pairs from the linear conversion
response, preserving the symmetry of the conversion and then recalculating the ellipticity angles of the resulting
eigenstates. The application of this approach is illustrated in Fig. \ref{ellipticities}(b) for the reference case.
One can see that at $\nu_1$ the removal of just $p_{x}$ and $m_{x}$ leads to a decrease of $\left | \epsilon \right
|$ to approximately $\pi/60$ for both eigenstates with their polarizations collapsing to almost linear states. Such
a dramatic transition towards purely anisotropic response confirms that on the microscopic level the effect of
circular dichroism at $\nu_1$ results from electric and magnetic dipolar charge-current excitations induced in the
metamolecules. Using the same approach for the combination of $T_{y}$ and $Q_{e}$ we also observed a clear change
in the polarization eigenstates but at $\nu_2$, where $\left | \epsilon \right |$ decreased to approximately $
\pi/30$ [see Fig. \ref{ellipticities}(c)]. The somewhat larger new values of $\left | \epsilon \right |$ here
compared to the reference case at $\nu_1$ are attributed to the non-negligible contribution of $p_{x}$ and $m_{x}$
that also impacts the optical activity at $\nu_2$, as evident from Fig. \ref{ellipticities}(b). Nevertheless, the
resulting profound modification of the eigenstates at $\nu_2$ supports our initial conclusion that the resonant
excitation of toroidal dipole and electric quadrupole is the dominant mechanism of the effect of circular dichroism
at this frequency.

The presence of strong toroidal and quadrupole contributions to the optical activity of the metamaterial may seem surprising. Nothing of the kind has been observed in chemistry, where the circular dichroism is a standard analytical technique that reports, for example, on secondary and tertiary structures of proteins \cite{Berova2000}. Indeed, the conventional expression for the rotatory power does not feature toroidal or quadrupolar moments \cite{Condon1937}:

\begin{equation}
R_{nk}=Im\left \lfloor \left \langle \psi _k\left | \mathbf{\hat{p}} \right | \psi_n\right \rangle \left \langle \psi _k\left | \mathbf{\hat{m}} \right | \psi_n\right \rangle\right \rfloor,
\label{TOAform1}
\end{equation}

and contains only the electric and magnetic dipole operators $\mathbf{\hat{p}}$ and $\mathbf{\hat{m}}$. This expression is obtained using perturbation theory under the common (in chemical sciences) assumption that the size of a molecule is much smaller that the wavelength of light (i.e. $\mathbf{k}\cdot \mathbf{r}\ll 1$, where $\mathbf{k}$ is the wavevector). Such assumption, however, is no longer valid for the metamolecule, and therefore the power series expansion of the complex exponential form of the plane wave must be carried out to at least the second order of $\mathbf{k}\cdot \mathbf{r}$. In particular, for the vector potential we have:

\begin{align}
\mathbf{A} &=\mathbf{A}_0Re\left \{ e^{-i\omega t} e^{i\mathbf{k}\cdot \mathbf{r}}\right \} \\
 & \approx \mathbf{A}_0Re\left \{ e^{-i\omega t}\left ( 1+i\mathbf{k}\cdot \mathbf{r}-\frac{\left ( \mathbf{k}\cdot \mathbf{r} \right )^2}{2} \right ) \right \}
 \label{TOAform2}
\end{align}

Because of the second order term $\left (\mathbf{k}\cdot \mathbf{r} \right )^2$ the magnetic perturbation operator $\mathbf{A}\cdot \mathbf{\hat{p}}$ \cite{Jensen2007} acquires the following components:

\begin{equation}
\left ( \mathbf{k}\cdot \mathbf{r} \right )^2 \vec{\nabla}=(\mathbf{k}\cdot\mathbf{k})(\mathbf{r}\cdot\mathbf{r})\vec{\nabla}-\left ( \left [ \mathbf{k}\times \mathbf{r} \right ]\cdot \left [ \mathbf{r}\times \mathbf{k} \right ] \right )\vec{\nabla},
\label{TOAform3}
\end{equation}

where,

\begin{equation}
(\mathbf{r}\cdot\mathbf{r})\vec{\nabla}=\mathbf{r}\left ( \mathbf{r}\cdot \vec{\nabla}\right )-\left \lfloor \mathbf{r} \times\left \lfloor   \mathbf{r} \times \vec{\nabla}\right \rfloor \right \rfloor
\label{TOAform4}
\end{equation}

The last term in Eq. \ref{TOAform4} corresponds to the toroidal moment. The presence of toroidal contributions in the circular dichroism is therefore the consequence of non-local response of the metamolecules. The complete derivation of the expression for the rotatory power that explicitly features the toroidal dipole operator $\mathbf{\hat{T}}$ will be published elsewhere.

In summary, we investigated experimentally and theoretically a structurally chiral metamaterial that exhibits
strong circular dichroism and supports non-negligible toroidal dipolar response. Analysis of the fields scattered
by the charge-current multipoles induced in the metamolecules and the polarization eigenstates of the system
reveals that the metamaterial exhibits both conventional optical activity, by way of electric and magnetic dipole
coupling, and, previously unobserved, strong optical activity underpinned by the resonant excitation of toroidal
dipole and electric quadrupole. Such an exotic multipole combination is anticipated to be the cause of (or at least
strongly contribute to) optical activity in many other material systems with toroidal topology, common in chemistry
and biology.

\begin{acknowledgments}
The financial support of DSTL, the Enginneering and Physical Sciences Research Council, UK and the Leverhulme Trust
is acknowledged. The authors would like to thank Vassili Savinov for numerous fruitful discussions.
\end{acknowledgments}

\end{document}